\documentclass[journal=apchd5,manuscript=article]{achemso}

\usepackage[version=3]{mhchem} 
\usepackage[T1]{fontenc}       
\usepackage{amsmath}
\usepackage{graphicx}
\usepackage{bm}
\usepackage{times}
\usepackage{ulem}
\usepackage{color}
\usepackage[euler]{textgreek}

\author{Seng Fatt Liew}
\affiliation[AP]{Department of Applied Physics, Yale University, New Haven, CT 06520, USA}
\author{S\'{e}bastien M. Popoff}
\affiliation[AP]{Department of Applied Physics, Yale University, New Haven, CT 06520, USA}
\alsoaffiliation{CNRS LTCI, Telecom ParisTech, 46 rue Barrault, 75013 Paris, France}
\author{Stafford W. Sheehan}
\affiliation[Chem]{Department of Chemistry, Yale University, New Haven, CT 06520, USA}
\author{Arthur Goetschy}
\affiliation[AP]{Department of Applied Physics, Yale University, New Haven, CT 06520, USA}
\alsoaffiliation{ESPCI ParisTech, PSL Research University, Institut Langevin, 1 rue Jussieu, 75005 Paris, France}
\author{Charles A. Schmuttenmaer}
\affiliation[Chem]{Department of Chemistry, Yale University, New Haven, CT 06520, USA}
\author{A. Douglas Stone}
\affiliation[AP]{Department of Applied Physics, Yale University, New Haven, CT 06520, USA}
\author{Hui Cao}
\affiliation[AP]{Department of Applied Physics, Yale University, New Haven, CT 06520, USA}
\email{hui.cao@yale.edu}

\title{Coherent control of photocurrent in a strongly scattering photoelectrochemical system}

\begin{document}

	

\begin{abstract}
  A fundamental issue that limits the efficiency of many photoelectrochemical systems is that the photon absorption length is typically much longer than the electron diffusion length. 
  Various photon management schemes have been developed to enhance light absorption; one simple approach is to use randomly scattering media to enable broadband and wide-angle enhancement. 
  However, such systems are often opaque, making it difficult to probe photo-induced processes. 
  Here we use wave interference effects to modify the spatial distribution of light inside a highly-scattering dye-sensitized solar cell to control photon absorption in a space-dependent manner. 
  By shaping the incident wavefront of a laser beam, we enhance or suppress photocurrent by increasing or decreasing light concentration on the front side of the mesoporous photoanode where the collection efficiency of photoelectrons is maximal. 
  Enhanced light absorption is achieved by reducing reflection through the open boundary of the photoanode via destructive interference,
  leading to a factor of two increase in photocurrent.  
  This approach opens the door to probing and manipulating photoelectrochemical processes in specific regions inside nominally opaque media. 
\end{abstract}
\section*{TOC Graphic}
\begin{scheme}
	\centering
	\includegraphics[width=0.5\linewidth]{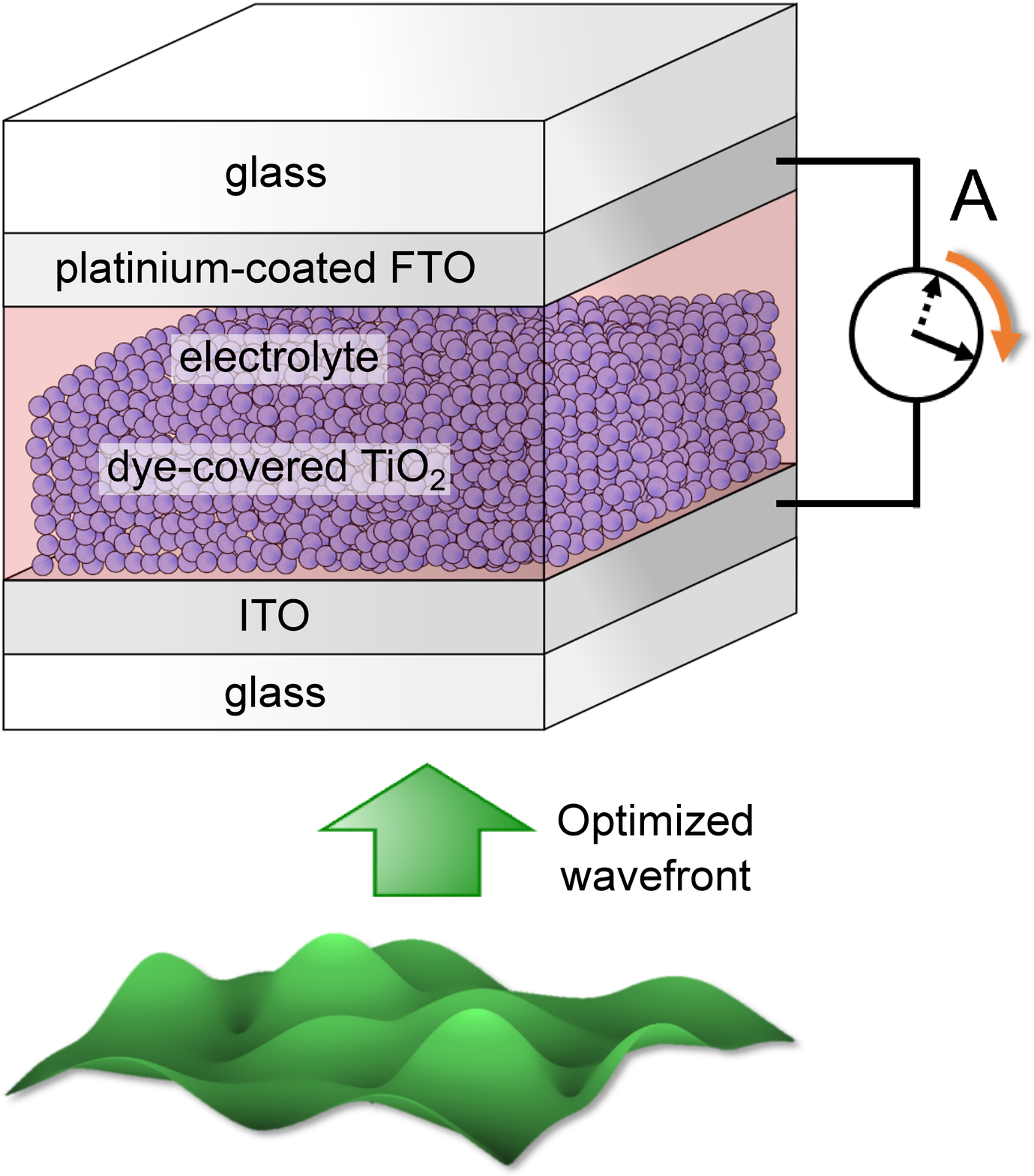}
	\caption{}
	\label{}
\end{scheme}

\section*{Introduction}
Optical scattering, induced by structural disorder, can increase the dwell time of light inside an absorbing material, thus enhancing the chance of absorption~\cite{Yablonovitch_Cody,Lederer_OE, Polman_NanoLett,Kimerling_OE,Wiersma_NatMat}. 
Compared to other light-trapping schemes that have been developed for solar cells, the non-resonant nature of this scheme enables broadband enhancement for light with a wide angle of incidence \cite{Atwater_NatMat,Peumans}. 
However, the degree of absorption does not depend solely on the absorption coefficient and scattering properties of the medium, but also on the coherence and spatial characteristics of the input light. 
One recent example confirming this is the demonstration of coherent perfect absorption, which is the time-reversed analog of laser emission~\cite{chong10CPA,wan2011time}. 
Similar to lasing, this effect is narrow-band and requires tuning of both the frequency of the input light and absorptivity of the material to achieve 100\% absorption~\cite{nohPRL12,nohOE13,liSciRep14}. 
A related but distinct approach, termed coherently enhanced absorption, takes advantage of the large number of degrees of freedom available for multiple scattering in a lossy environment to control the absorption~\cite{chong2011hidden,Goetschy2013}. 
It has been shown theoretically an appropriate coherent superposition of input fields can significantly enhance the coupling of light into a scattering medium, increasing the absorption at any frequency over a broad range. 

Recent advances in wavefront shaping (WFS) techniques using spatial light modulators (SLMs) have enabled the generation of arbitrary wavefronts at the input of an optical system. 
This technique is driving forward the development of optical imaging techniques through scattering media such as biological tissue and fog~\cite{YangNatPho08,Popoff2010a,Mosk2012,Bertolotti2012}, and also enables the control of light intensity~\cite{YangNatPho08,vellekoop2007focusing,vellekoop2008demixing,vellekoop2010exploiting}, polarization state~\cite{Guan2012control,Park2012active},  spectrum~\cite{Park2012spectral,Small2012control}, or temporal profile~\cite{katz2011focusing,AulbachPRL11,AulbachOE12,ShiOL13} at the scale of wavelength (single or a few speckles). 
While the wavelength-scale control is desired for imaging with high resolution \cite{JudkewitzNatPho13}, the ability to manipulate light distributions on a much larger scale in a disordered system is needed for many applications. 
One example is photoelectrochemical cells, which often utilize light scattering to increase photon dwell time to enhance absorption. 
As the photoelectrons are generated in a volume that contains hundreds of thousands of speckles, increasing the intensity of a single speckle even by a few thousand times makes little difference to the total photocurrent. 
Hence,  global control of the light distribution is necessary, but this is known to be a much more challenging optimization problem than local control. 
Good progress has been made recently to enhance the total transmission through a lossless opaque medium via WFS of an incident laser beam~\cite{Goetschy2013,vellekoopPRL08,Kim2012,Yu2013,popoff2013coherent}. 
However, coherent control of light distribution and absorption over a macroscale region in highly scattering media has never been demonstrated. 

Here we use WFS to both enhance and suppress the photocurrent in a dye-sensitized solar cell (DSSC), with a photoanode comprised of a highly scattering layer of dye-sensitized TiO$_2$ nanoparticles. 
Since the collection efficiency of charge carriers is highly dependent on where light is absorbed~\cite{gratzel01}, it is advantageous to trap light inside the scattering layer as close as possible to the interface with the electrode, where charges can be efficiently separated, thus minimizing charge recombination~\cite{lin11}. 
However, strong light leakage through the interface makes it extremely difficult to confine light to a region close to the open boundary and enhance the absorption there. 
But here we show a significant enhancement of photocurrent by optimizing the wavefront of an incident laser beam, and illustrate the underlying mechanism via numerical modeling. 
In addition, we show that WFS can also {\it suppress} absorption by increasing the amount of light reflected. 
This work demonstrates the great potential of WFS toward depositing energy into targeted regions inside turbid media, which will enable the spatially resolved characterization of light absorption and charge transport inside regenerative and artificial photosynthetic systems. 

\section*{Control of DSSC photocurrent}
\begin{figure}
	\centering
	\includegraphics[width=\textwidth]{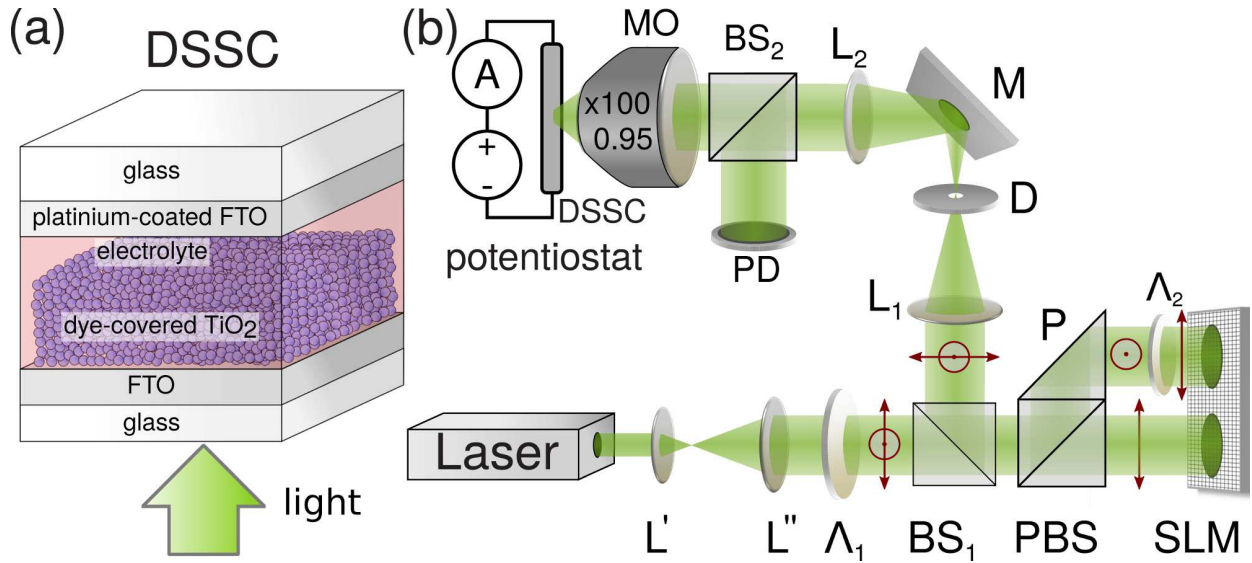}
	\vspace{1cm}
	\caption{Experiment setup.  
		(a) Schematic of the dye-sensitized solar cell (DSSC) used in this study, which employs a thinner absorbing layer and a different electrolyte from the conventional DSSC to demonstrate coherent control of photovoltaic current. (b) Schematic of the experimental setup (see Methods). A laser beam at $\lambda = 532$nm is split to two orthogonal polarizations and projected onto the SLM. The phase front of both polarizations are modulated and then recombined at a beam splitter. The shaped wavefront is imaged onto the surface of the porous electrode in the DSSC using a microscope objective. The illuminated area is a disk of diameter $\sim 8.3$ $\mu$m. The short-circuit current of the solar cell is measured using a potentiostat. 
	}
	\label{Setup}
\end{figure}	 
A DSSC comprises a porous layer of sintered TiO$_2$ nanoparticles covered by a molecular dye, in this case the ruthenium-based N719 dye~\cite{ORegan1991}, deposited on a conductive glass substrate. 
It functions as the light-absorbing working electrode of the solar cell, while a platinum-loaded conductive glass slide acts as the counter electrode. 
An electrolyte fills the gap between the two electrodes and shuttles charges between them. 
The working electrode of a DSSC constructed in this manner can be considered as a scattering sample with homogeneous absorption~\cite{MiguezEES2014}.
Figure~\ref{Setup}a presents a schematic of the solar cells created in this study. 
Light is incident onto the porous TiO$_2$ layer through the glass substrate coated with transparent conducting oxide. 
After light absorption, the excited dye molecules rapidly inject electrons into the TiO$_2$. 
These electrons diffuse through the sintered particle network to be collected at the front side electrode, while the dye is regenerated via reduction by a redox mediator in the electrolyte, in this case the iodide/triiodide redox couple. 
The photo-induced current from a DSSC is typically measured without applying a bias voltage, and is known as the short-circuit current $I_{sc}$. 
Not all photo-generated electrons can diffuse through the TiO$_2$ layer and reach the electrode, because electrons may recombine or get trapped on the way~\cite{NakadeJPCB02, OekermannJPCB04}. 
The farther the electrons are generated from the electrode, the lower their probability of reaching it and contributing to the current.   
Hence, $I_{sc}$ depends not only on the amount of light being absorbed but also on the position where the light is absorbed inside the TiO$_2$ layer~\cite{StaffJPCC2013}. 

In the WFS experiment, the solar cell is excited by a narrow band laser beam at wavelength $\lambda$ = 532 nm. 
The input wavefront is modulated using the experimental setup schematically shown in Fig.~\ref{Setup}b. 
The two linear polarizations are modulated independently~\cite{popoff2013coherent} (see Methods) and recombined into a single beam, which is projected onto the scattering medium in the solar cell using a microscope objective (NA = 0.95). 
We measure $I_{sc}$ generated by the DSSC for each input phase mask and use a genetic algorithm to find the wavefronts that maximize or minimize $I_{sc}$. 
The genetic algorithm is known to be robust against noise in WFS experiments~\cite{katz2011focusing,conkey2012genetic}. 
A cost function $\eta_e$, which estimates the variation of the conversion efficiency of the solar cell, is defined as:
\begin{equation}
\eta_e = \frac{(I_{sc}-I_{dark})/\mathcal{P}_{0}}{(\left\langle I_{sc} \right\rangle-I_{dark})/\left\langle \mathcal{P}_{0} \right\rangle},
\end{equation}
\noindent where $\left\langle I_{sc} \right\rangle$ and $\left\langle \mathcal{P}_{0} \right\rangle$ are the average short-circuit current and input optical power, respectively, measured for a set of random phase masks. 
$I_{dark}$ is the small dark current measured at $\mathcal{P}_{0} = 0 \mu$W ({\it SI}). 

\begin{figure}
	\centering
	\includegraphics[width=\linewidth]{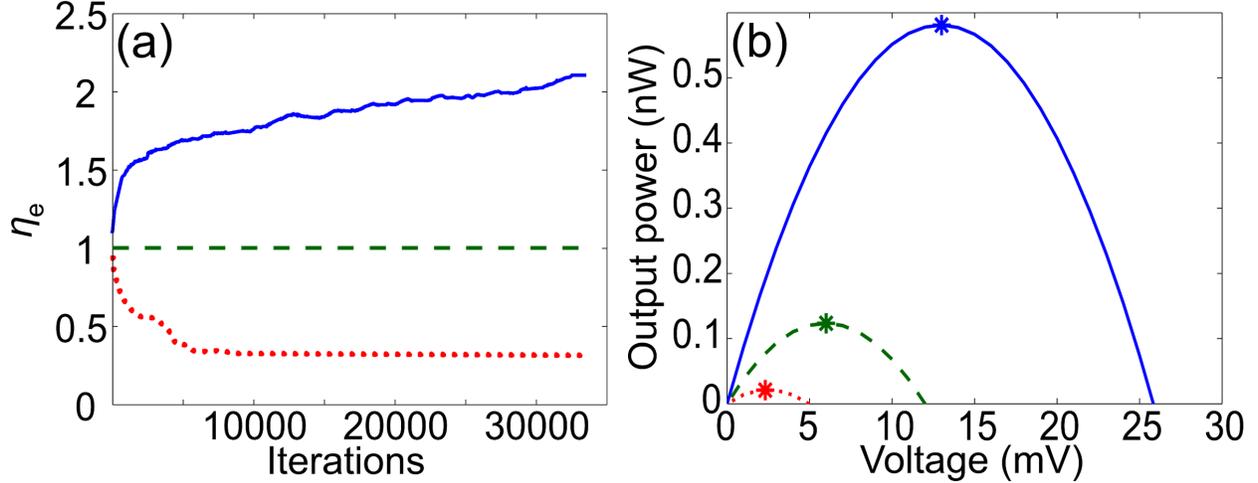}
	\caption{Control of the DSSC current and power by shaping the input wavefront of a laser beam. 
		(a) The measured cost function $\eta_e$ for the short-circuit current, normalized to the value with random input wavefront (green dashed line), increases (blue solid line) or decreases (red dotted line) during the optimization of the input wavefront by a genetic algorithm. The photo-induced current is enhanced $2.1$ times or suppressed $3.0$ times by wavefront shaping. 
		(b) Output electrical power of the DSSC illuminated by a random phase pattern (dashed green line), and the two optimized wavefronts for maximal $\eta_e$ (solid blue line) and minimal $\eta_e$ (dotted red line). 
		The incident laser power is fixed at $100 \mu$W. The asterisk indicates the maximal power, which is enhanced 4.7 times or reduced 5.7 times. 
	}
	\label{results1}
\end{figure}	

We constructed the DSSC using standard procedures (see Methods and {\it SI}). 
The transport mean free path  of light in the  $20 \pm 1$ $\mu$m thick TiO$_2$ layer, $\ell$, is measured in a coherent backscattering experiment to be 3.2 $\mu$m  \cite{akkermans2007mesoscopic}. 
To shorten the response time of the solar cell in order to perform WFS experiments on reasonable time scales, we use a concentrated non-volatile iodine-based electrolyte that also absorbs light at 532 nm. 
The absorption of light by the electrolyte is proportional to that by the dye at any location, as both scale linearly with local field intensity. 
Thus the short-circuit current $I_{sc}$, which results from light absorption by the dye, also reflects the absorption of the electrolyte, which does not contribute directly to the current. 
The ballistic absorption length $\ell_a$ is about 65 $\mu$m, and the diffusive absorption length (the typical distance light will travel via multiple scattering prior to being absorbed), $L_a=\sqrt{\ell \ell_a/3}$, is 8.4 $\mu$m~\cite{akkermans2007mesoscopic,van_rossum}. 
Without absorption, the total transmission $T$ of the TiO$_2$ layer is estimated to be $26\%$, and the reflection $R = 74\%$. 
In the presence of absorption, $T$ is reduced to $7\%$, and $R$ to $47\%$, with the remaining $46\%$ being absorbed (in the absence of wavefront optimization). 
These estimations agree well with the experimental measurement ({\it SI}). 

By optimizing the incident wavefront, the absorption cost function $\eta_e$ is gradually increased or decreased as shown in Fig.~\ref{results1}a. 
With the optimal wavefronts, the photocurrent is enhanced by a factor of $\sim 2.1$ or suppressed by a factor of $\sim 3.0$. 
We are thus able to coherently vary the energy conversion efficiency by a ratio of $\sim 6.3$. 
To quantify the variation of the output power due to the modification of the input optical pattern, we measure the current-voltage ($I\text{-}V$) curves for the two optimized wavefronts corresponding to maximum and minimum $\eta_e$, as well as for a random illumination pattern. 
Figure~\ref{results1}b displays the light-induced electrical output power, given by $\left(I-I_{dark}\right) \, V$, as a function of bias voltage for each scenario. 
The maximum output power is enhanced 4.7 times or reduced 5.7 times, leading to total variation by a factor of 27. 

Since approximately $46\%$ of the light is absorbed in the solar cell for a random illumination, the highest possible enhancement of the absorption would be a factor of 2.2. 
If $I_{sc}$ were proportional to the total amount of light being absorbed~\cite{StaffJPCC2013}, the experimental enhancement of 2.1 of the short circuit current would correspond to an almost perfect absorption of light in the absorbing medium. 
However, achieving the maximum absorption requires complete control over the amplitude and phase of the light in all input channels~\cite{chong2011hidden}. 
Experimentally we control only a limited fraction of the input channels due to the finite numerical aperture of the objective lens and the fact that the probe beam is incident only on one side of the scattering sample. 
Moreover, our SLM setup only modulates the phase and not the amplitude of the incident light, as would be necessary to achieve the global optimum. 
Consequently, the enhancement of total absorption that could be achieved with WFS is significantly reduced~\cite{Goetschy2013, popoff2013coherent}. 

\section*{Manipulation of local absorption}
To investigate the cause for such a high enhancement factor measured experimentally, we perform numerical modeling. 
Without loss of generality, we simulate the scattering medium confined in a 2D waveguide with reflecting sidewalls to reduce the computation load; the system dimensions, as well as the amount of scattering and uniform absorption, are chosen to be near to the experimental values (see Methods). 
Light with transverse-magnetic polarization ($E$ field perpendicular to the waveguide plane) is injected from one end of the waveguide, and its propagation in the random system is calculated with the recursive Green's function method~\cite{chong2011hidden,leePRL81,beenakkerPRL96}. 
The transmission and reflection of light for the systems with and without absorption, averaged over random input wavefronts, match the measured values of real samples. 
To model the effect of WFS in our experiment, we use the empty waveguide modes as the basis to shape the input field. 
During the optimization process, we modulate the phase of each empty waveguide modes while keeping their amplitude constant. 
The (linear) space-dependent absorption of light is proportional to the local intensity $|E(x,y)|^2$, and the total absorption is proportional to the spatially integrated intensity. 
To take into account the spatial variation of the electron collection efficiency $D(x) = \exp[-x/\xi]$, where $\xi$ is the electron diffusion length, we define a new cost function $\eta_s$ which models the relative change of the short-circuit current similarly to $\eta_e$:
\begin{equation}
\eta_s = \frac{\int\int D(x) |E(x,y)|^2 dx \, dy}{\langle\int\int D(x) |E_0(x,y)|^2 dx \, dy\rangle},
\end{equation}
where $E(x,y)$ is the electric field distribution in the scattering medium for the optimized input wavefront, $E_0(x,y)$ is the field distribution for a random input wavefront, and $\langle .. \rangle$ represents the average over random input wavefronts. 
As in the solar cell, the absorption of light that occurs far from the front surface of the random medium contributes little to $\eta_s$. 
We used both a genetic algorithm and a sequential algorithm \cite{vellekoop2007focusing,VellekoopSequential08} to maximize or minimize $\eta_s$, and found similar results, although the sequential algorithm converges faster than the genetic algorithm due to the absence of noise in simulation. 

\begin{figure}
	\centering
	\includegraphics[width=\linewidth]{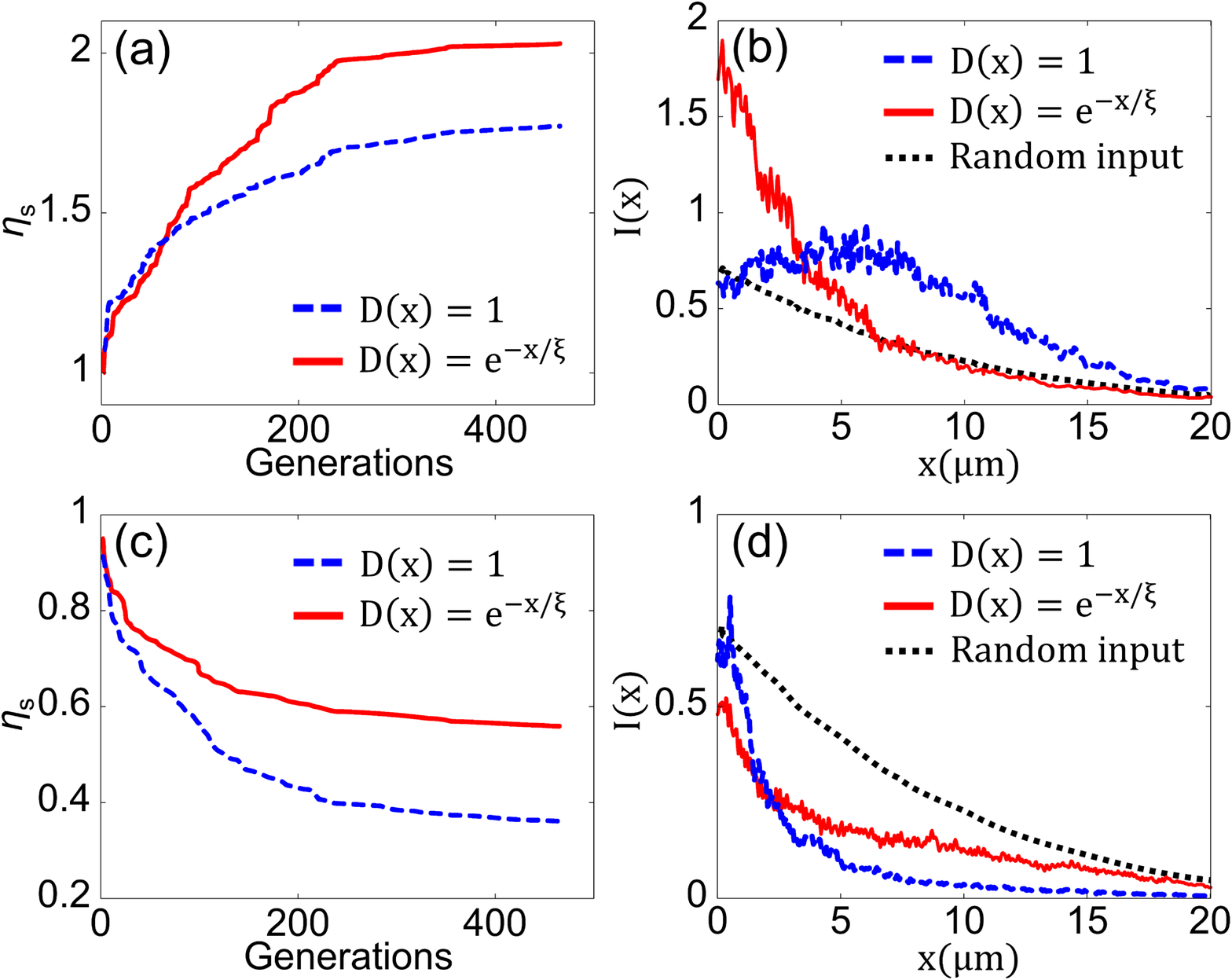}
	\caption{
		Numerical simulation of WFS experiment. 
		(a) The cost function $\eta_s$, defined in Eq.(2), increases and then saturates when the input wavefront is optimized for $D(x) = \exp[-x/\xi]$ (red solid line) and $D(x) = 1$ (blue dashed line) using a sequential algorithm.  
		$\xi$ is taken to be 3 $\mu$m (see Methods). 
		(b) The cross-section integrated intensity in the random system illuminated by random input wavefronts (black dotted line) and the optimized wavefront obtained with the non-uniform cost function for $D(x) = \exp[-x/\xi]$ (red solid line) and the uniform cost function for $D(x) = 1$ (blue dashed line). 
		The light is concentrated near the front side of the random system for $D(x) = \exp[-x/\xi]$, but penetrates deeper into the system for $D(x) = 1$.
		(c,d) Same as (a,b) for the minimization of $\eta_s$. 
		For $D(x) = \exp[-x/\xi]$, the light intensity near the front side of the random system is reduced, but increased deeper into the system. 
		For $D(x) = 1$, the light intensity is reduced in the interior of the system but remains almost the same near the front surface. 
	}
	\label{RGFsim}
\end{figure}	 

Figure~\ref{RGFsim}a shows the enhancement of the cost function $\eta_s$ in the random medium with a uniform absorption coefficient. 
For comparison we also simulate the case of $D(x) = 1$, which corresponds to a uniform cost function similar to the standard coherent enhancement of global absorption~\cite{chong2011hidden,Goetschy2013}. 
Both the non-uniform ($D(x) = \exp[-x/\xi]$) and uniform cost functions increase continuously in the optimization process and eventually saturate; however as expected, the final enhancement factor for $D(x) = \exp[-x/\xi]$  significantly exceeds that for $D(x) = 1$.
Figure \ref{RGFsim}b represents the integrated intensity $I(x) = \int |E(x,y)|^2 dy$ inside the random medium. 
With random wavefront illumination, $I(x)$ decays exponentially in the disordered system, as expected for light diffusion in the presence of absorption \cite{van_rossum}. 
When the input wavefront is optimized for uniform cost function ($D(x) = 1$), $I(x)$ is peaked in the interior of the random system. 
The buildup of energy deep inside the scattering medium enhances the total absorption from $46\%$ to $82\%$. 
Meanwhile, the reflection decreases from $47\%$ to $9\%$, and the transmission slightly increases from $7\%$ to $9\%$. 
This result confirms that the global absorption is enhanced by WFS through causing the light to penetrate deeper into the lossy disordered medium. 

In contrast, the optimization for the non-uniform cost function ($D(x) = \exp[-x/\xi]$) leads to a significant increase of light intensity close to the front surface of the random medium [Fig. \ref{RGFsim}b], since only this region contributes significantly to the cost function $\eta_s$. 
Naively, the proximity to the open boundary would be expected to facilitate light escape from the absorbing medium, thus reducing the absorption and increasing the reflection. 
Instead, for the optimal wavefront, the total absorption is increased from $46\%$ to $67\%$, and the reflection decreased from $47\%$ to $28\%$. 
This result suggests some partial trapping of the light near the surface, leading to a buildup of its intensity in the vicinity of the open boundary where the front side electrode is located. 
Hence, the numerical simulation illustrates that our experimental procedure, which does feedback optimization based on the short-circuit current, does not simply optimize total absorption, but optimizes absorption locally, near the working electrode, so as to preferentially enhance the collected current.   
The measured current increase by a factor of 2.1 does not imply perfect absorption of all incident light and can be realized with the incomplete experimental control. 

\section*{Modal interference effects}
To understand how the leakage rate of the light close to the open boundary is reduced, we investigate the contribution of different quasi-modes in the system to the reflected light. 
It has been shown previously that  light propagation in random media can be well explained in terms of  the interference effects of quasi-modes \cite{genack_nature, genack_JMP, liew_PRB, genack_arxiv14}. 
We take this approach by decomposing the electric field inside the random medium in terms of the quasi-modes of the system:
\begin{equation}
E(x,y) = \sum_m a_m u_m(x,y) + \sum_m b_m v_m(x,y),
\end{equation}
\noindent where $u_m$ ($v_m$) is the $m$-th mode with purely outgoing (incoming) wave boundary condition, and $u_m = v_m^*$~\cite{I-SALT}. 
The reflected field at the front surface ($x=0$) of the disordered medium can be expressed as $E_r(y) = \sum_m a_m u_m(0, y)$. 
The total intensity of reflected waves is  $R = \int |E_r(y)|^2 dy = \sum_m \alpha_m$, where $\alpha_m = a_m \int E_r^*(y) u_m(0,y) dy$ represents the contribution of the $m$-th mode to the reflection.
While $\alpha_m$ is a complex number, $R$ is a real number, and the contributions of different modes to $R$ interfere with each other.  
The interference effect can be quantified by $C \equiv |\sum_m \alpha_m|^2/\sum_m |\alpha_m|^2$ ({\it SI}). 
If $C>1$ ($C<1$), the modes interfere constructively (destructively) at the front surface to enhance (suppress) reflection. 

\begin{figure}
	\centering
	\includegraphics[width=\linewidth]{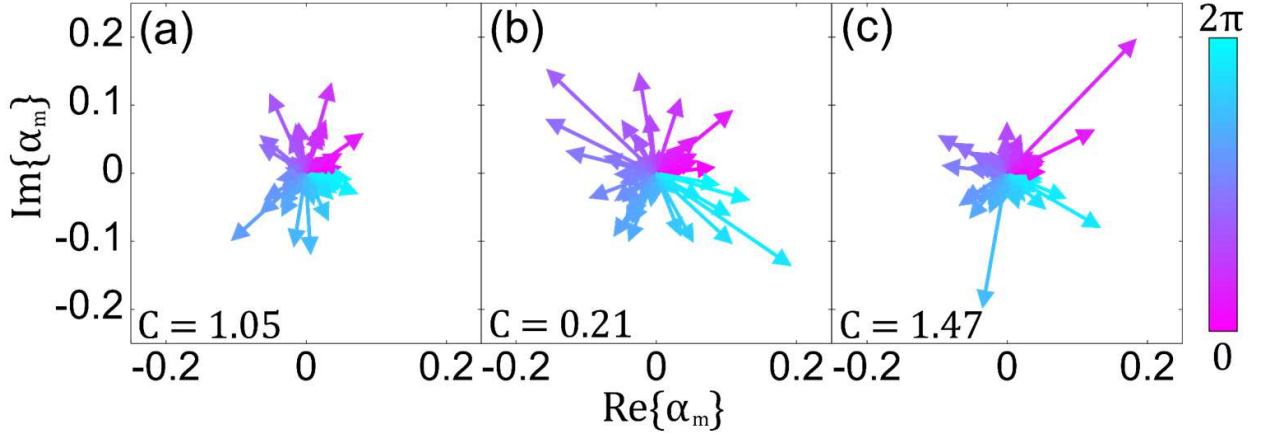}
	\caption{
		Modal analysis of the wavefront optimization. 
		Contributions $\alpha_m$ of individual modes $m$ supported by the random system to the total intensity of reflected waves. The modes are excited by a random input wavefront (a), the optimized wavefront for maximal $\eta_s$ (b) and minimal $\eta_s$ (c) with $D(x) = \exp[-x/\xi]$. In (a) the modes have random phases and their interference is averaged out, $C =1.05$. In (b) the modes interfere destructively, $C=0.21 < 1$, to reduce reflection and enhance absorption. In (c) the modes interfere constructively, $C = 1.47 > 1$, to enhance reflection and reduce absorption. 
	}
	\label{modepic}
\end{figure}
To reduce the computing time for the full modal analysis, the length and width of the system in the simulation is reduced by a factor of two. 
The scattering and absorption parameters are adjusted so that the total transmission, reflection and absorption for random input wavefronts are essentially unchanged. 
We use the finite-difference frequency-domain (FDFD) method to calculate the quasi-modes whose frequencies are close to the frequency of the input light. 
The calculated values of $\alpha_m$ for the different scenarios with $D(x) = \exp[-x/\xi]$ are plotted in the complex plane in Figs. \ref{modepic}a-c. 
The modes have uncorrelated phases for random wavefront illumination and their interference is averaged out (and therefore $C$ is close to 1). 
We see that the optimized wavefront for enhanced local absorption causes a destructive interference among the modes, making $C$ much smaller than 1, and more importantly, that reflection is minimized and absorption is maximized at the front surface.
Despite the complexity of the medium, WFS exploits intereferences to efficiently concentrate light in the targeted region of the system.

Similarly, we show that WFS can also reduce the energy in a chosen region by minimizing the cost function $\eta_s$ [Figs. \ref{RGFsim}c and d].
For $D(x) = 1$, the reflection is greatly enhanced, decreasing the light intensity deep inside the system and thus the total amount of absorption.
For $D(x) = \exp[-x/\xi]$, the minimization of the cost function is less efficient as it requires us to {\it simultaneously} increase the reflection and decrease the absorption close to the front surface.
By decomposing the fields into the quasi-modes of the system, we confirm that the increase in reflection is caused by the constructive interference of excited modes [Fig. \ref{modepic}c], which facilitates light escape and suppresses absorption. 

\section*{Conclusion}
In summary, we have experimentally demonstrated coherent control of light distribution and absorption over a length scale much larger than the wavelength of light inside an opaque mesoporous photoelectrochemical system that utilizes multiple scattering to improve light confinement. 
Our proof-of-concept experiment shows that it is possible to vary the photocurrent in a dye-sensitized solar cell (DSSC) by tailoring the input wavefront of a laser beam. 
To maximize the DSSC's short-circuit current, the optimal wavefront concentrates the input light to the front side of the porous photoanode, where photo-generated electrons are most efficiently collected. 
Despite the proximity to an open boundary, we find that destructive interference among the excited modes minimizes light leakage and enhances absorption. 

Although in the current WFS experiment a narrow-band laser is used for coherent control of absorption, a recent theoretical study suggests that this approach is applicable to a relatively broadband light~\cite{Hsu15}.
More generally, our method, assisted by a guidestar \cite{YangNatPho15}, will allow for spatially-resolved characterization of photo-induced processes inside opaque media, such as photoelectrochemical cells. 
This approach opens the possibility of investigating photoelectrochemical processes in targeted regions inside dense scattering media, such as that used in regenerative and artificial photosynthetic solar cells. 
Those studies will provide physical insight into the operation of photovoltaic and photoelectrochemical systems, potentially leading to new methods to improve photoconversion performance.   

\section*{Methods}
\textbf{Sample preparation.} 
To facilitate the WFS experiment, we made a DSSC with a thicker TiO$_2$ scattering layer, reduced loading of N719 dye, and a thin (175 $\mu$m thick) tin-doped indium oxide (ITO) coated glass substrate for the working electrode.
The porous TiO$_2$ working electrode is made from $\sim 20$nm TiO$_2$ nanoparticle paste (P25, Degussa).
The paste was doctor-bladed onto a 175 $\mu$m thick ITO-coated glass coverslip (SPI, Product No. 06477B-AB) 5 times. 
The sample was heated to 200$^{\circ}$C for 10 min then cooled to room temperature between each application of a layer of paste. 
The final thickness of the TiO$_2$ layer is $20 \pm 1$ $\mu$m. 
The working electrode is sintered at 450$^{\circ}$C for 1 hour, then immersed in a solution of N719 (ditetrabutylammonium cis-bis(isothiocyanato)bis(2,2'-bipyridyl-4,4'-dicarboxylato) ruthenium(II) for 30 minutes~\cite{ORegan1991}. 
The counter electrode is made by heating hexachloroplatinic acid in ethanol (2 mg/mL) to 400$^{\circ}$C for 15 minutes on a fluorine-doped tin oxide (FTO) coated glass slide. 
The solar cells are assembled with a 25 $\mu$m Surlyn spacer hot-pressed between the electrodes and filled with a non-volatile iodine-based electrolyte (Mosalyte TDE-250, Solaronix SA, Switzerland).

\textbf{Optical setup.}
A shown schematically in Fig. \ref{Setup}, the output beam from a continuous-wave Nd:YAG laser is expanded (lenses L' and L'') and projected onto a phase-only SLM  (Hamamatsu, X10468-01). 
A half-wave-plate ($\Lambda_1$) is used to balance the light intensities of two orthogonal linear polarizations. 
To control independently the two polarizations, a polarizing beam splitter (PBS) attached to a right-angle prism (P) separates the two polarizations of the laser beam, which are modulated by different areas of the SLM. 
The cube and the prism are mounted in the same holder to eliminate fluctuations from independent reflective elements in the two paths. A second half-wave-plate ($\Lambda_2$) rotates the polarization to match that of the SLM. 
The two illuminated areas on the SLM corresponding to the two polarizations of the input light are divided into $1740$ macropixels. 
The modulated wavefronts are then recombined and projected onto the pupil of a microscope objective (MO) with numerical aperture $0.95$. 
A beam splitter (BS$_2$) along with a photo-detector (PD) are used to measure the optical power $\mathcal{P}_{0}$ impinging on the solar cell. 
The photo-induced current of the DSSC is measured by a potentiostat (Digi-Ivy DY2116b). 

\textbf{Numerical simulation.} 
In the 2D model, dielectric cylinders with radius 50 nm and refractive index 2.5 are randomly positioned inside a waveguide of width $W = 20$ $\mu$m and refractive index 1.5. 
The length of the random structure is $L = 20$ $\mu$m, and the effective index of refraction is 1.9.
The wavelength of input light is $\lambda = 532$ nm, and the number of transverse guided modes in the waveguide is 114. 
The transport mean free path $\ell$ is $3.7$ $\mu$m.  
Without absorption and wavefront optimization, the transmission is $T = 26\%$  and the reflection is $R$ = $74\%$. 
A uniform absorption coefficient is introduced via the imaginary part of the refractive index to the random system. 
The diffusive absorption length is $8.5$ $\mu$m. 
$T$ is reduced to $7\%$, and reflection to $47\%$, thus the absorption $A = 46\%$.
These numbers match the measured values of our DSSC. 
The electron diffusion length $\xi$ is chosen to be $3$ $\mu$m, close to the value of DSSCs similar to ours~\cite{NakadeJPCB02, OekermannJPCB04}. 

In the modal analysis to decompose the field distribution in the random medium with the quasi-modes, we keep the wavelength $\lambda$ the same, but reduce the number of waveguide modes to 37. 
$\ell$ is reduced to 2 $\mu$m, and the diffusive absorption length to 4.2 $\mu$m. 
Without absorption, $T = 24\%$ and $R = 76\%$; with absorption, $T = 7\%$, $R = 46\%$, and $A = 47\% $.   
The electron diffusion length $\xi$ is also shortened to 1.5 $\mu$m. 
\begin{suppinfo}
More details on the synthesis and optical characterization of the DSSC. 
Additional numerical simulation results of WFS to maximize and minimize the cost function $\eta_s$. 
Detailed description of modal analysis on the electric field distribution inside the random medium.
\end{suppinfo}

\section*{Author Information}
S. F. Liew and S. M. Popoff contributed equally to this work. 

\begin{acknowledgement}
	
	We thank Chia Wei Hsu for stimulating discussions. This work is supported by the NSF under Grant No. ECCS-1068642. S.W.S. acknowledges the support of an NSF Graduate Research Fellowship under DGE 1122492. 
	
\end{acknowledgement}


\end{document}